\PassOptionsToPackage{table}{xcolor}

\documentclass[sigconf]{acmart}

\usepackage{tikz}
\usepackage[ruled,vlined]{algorithm2e}
\usepackage[most]{tcolorbox}
\usepackage{mathtools}
\usepackage{graphicx}
\usepackage{booktabs}
\usepackage{multirow}
\usepackage{array}
\usepackage{makecell}

\definecolor{blue_1}{RGB}{240,245,254}
\definecolor{gary_1}{RGB}{239,239,239}
\definecolor{pink_1}{RGB}{250,239,239}

\AtBeginDocument{%
  }

\copyrightyear{2026}
\acmYear{2026}
\setcopyright{cc}
\setcctype{by}
\acmConference[MM '26]{Proceedings of the 34th ACM International Conference on Multimedia}{November 10--14, 2026}{Rio de Janeiro, Brazil}
\acmBooktitle{Proceedings of the 34th ACM International Conference on Multimedia (MM '26), November 10--14, 2026, Rio de Janeiro, Brazil}
\acmDOI{10.1145/3767308.3835562}
\acmISBN{979-8-4007-2213-4/2026/11}

\begin{document}

%%
%% Title
%%
\title[AuEmoChat]{%
  AuEmoChat: Authentic Emotion Understanding and Rendering
  for Conversational Speech Synthesis
}

%%
%% Authors and affiliations
%%
\author{Zhenqi Jia}
\email{jiazhenqi7@163.com}
\affiliation{%
  \institution{Inner Mongolia University}
  \city{Hohhot}
  \country{China}
}

\author{Yuan Zhao}
\email{zy404nf@163.com}
\affiliation{%
  \institution{Inner Mongolia University}
  \city{Hohhot}
  \country{China}
}

\author{Aruukhan}
\email{aruukhanres@gmail.com}
\affiliation{%
  \institution{Inner Mongolia University}
  \city{Hohhot}
  \country{China}
}

\author{Rui Liu}
\correspondingauthor
\email{imucslr@imu.edu.cn}
\affiliation{%
  \institution{Inner Mongolia University}
  \city{Hohhot}
  \country{China}
}

\author{Haizhou Li}
\email{haizhouli@cuhk.edu.cn}
\affiliation{%
  % \department{\mbox{SRIBD, School of Artificial Intelligence}}
  \institution{The Chinese University of Hong Kong}
  \city{Shenzhen}
  \country{China}
}

%%
%% Short author list used in page headers.
%%
\renewcommand{\shortauthors}{Jia et al.}

%%
%% Abstract
%%
\begin{abstract}
Conversational Speech Synthesis (CSS) aims to synthesize speech with human-like emotional expression and contextual consistency in user-agent interactions. Existing CSS methods struggle to render authentic human emotions due to limited predefined emotion label spaces (e.g., seven emotion categories), while redundant multimodal tokens in multi-turn dialogue history interfere with context understanding. To address these issues, we propose \textbf{AuEmoChat}, a CSS framework for authentic emotion understanding and rendering. First, we develop \textit{AuEmoCodec}, which learns a discrete authentic emotion token space from large-scale emotional speech via finite scalar quantization, enabling a more authentic emotion representation than limited basic emotion categories. Furthermore, we propose \textit{AuEmoToMe}, an authentic-emotion-guided token merging algorithm that merges redundant tokens in multimodal dialogue history while preserving emotion-relevant context. We integrate it into an autoregressive text-speech model to predict the target authentic emotion token and speech tokens. Finally, we propose \textit{Authentic Emotion Flow Matching}, which renders speech by jointly conditioning on merged dialogue context, target authentic emotion, and acoustic priors. Extensive experiments on the NCSSD-EmCap dataset demonstrate that AuEmoChat outperforms state-of-the-art CSS baselines and generates more expressive and authentic emotional speech. The code and speech demos will be available at: https://github.com/AI-S2-Lab/AuEmoChat.
\end{abstract}

%%
%% ACM Computing Classification System
%%


%%
%% Keywords
%%

%%
%% Uncomment and replace this block only when using a teaser figure.
%%
% \begin{teaserfigure}
%   \centering
%   \includegraphics[width=\textwidth]{figures/teaser.pdf}
%   \caption{Overview of the proposed AuEmoChat framework.}
%   \Description{An overview of AuEmoChat, including AuEmoCodec,
%   AuEmoToMe, autoregressive token prediction, and authentic emotion
%   flow matching.}
%   \label{fig:teaser}
% \end{teaserfigure}

%%
%% Generate title, authors, affiliations, ACM reference format,
%% CCS concepts, keywords, and copyright information.
%%
\maketitle

\section{Introduction}
Conversational Speech Synthesis (CSS) aims to leverage multimodal dialogue history to synthesize target speech with contextually appropriate affective prosody in User-Agent Interactions (UAI) \cite{guo2021conversational, liu2024emotion, hu-etal-2025-chain, leng2025eijl}. In UAI systems, the ability of the agent to generate speech that is appropriate for the dialogue context and conveys human-like emotional expression is vital for enhancing user experience. As UAI becomes increasingly prevalent, CSS has become a crucial component of intelligent interactive systems \cite{zhou2020design, seaborn2021voice, mctear2022conversational, zhu2025preference} and plays an important role in embodied intelligence applications such as virtual assistants, voice agents, intelligent robots, smart home devices, and in-vehicle infotainment systems.

\begin{figure}
    \centering
    \includegraphics[width=1\linewidth]{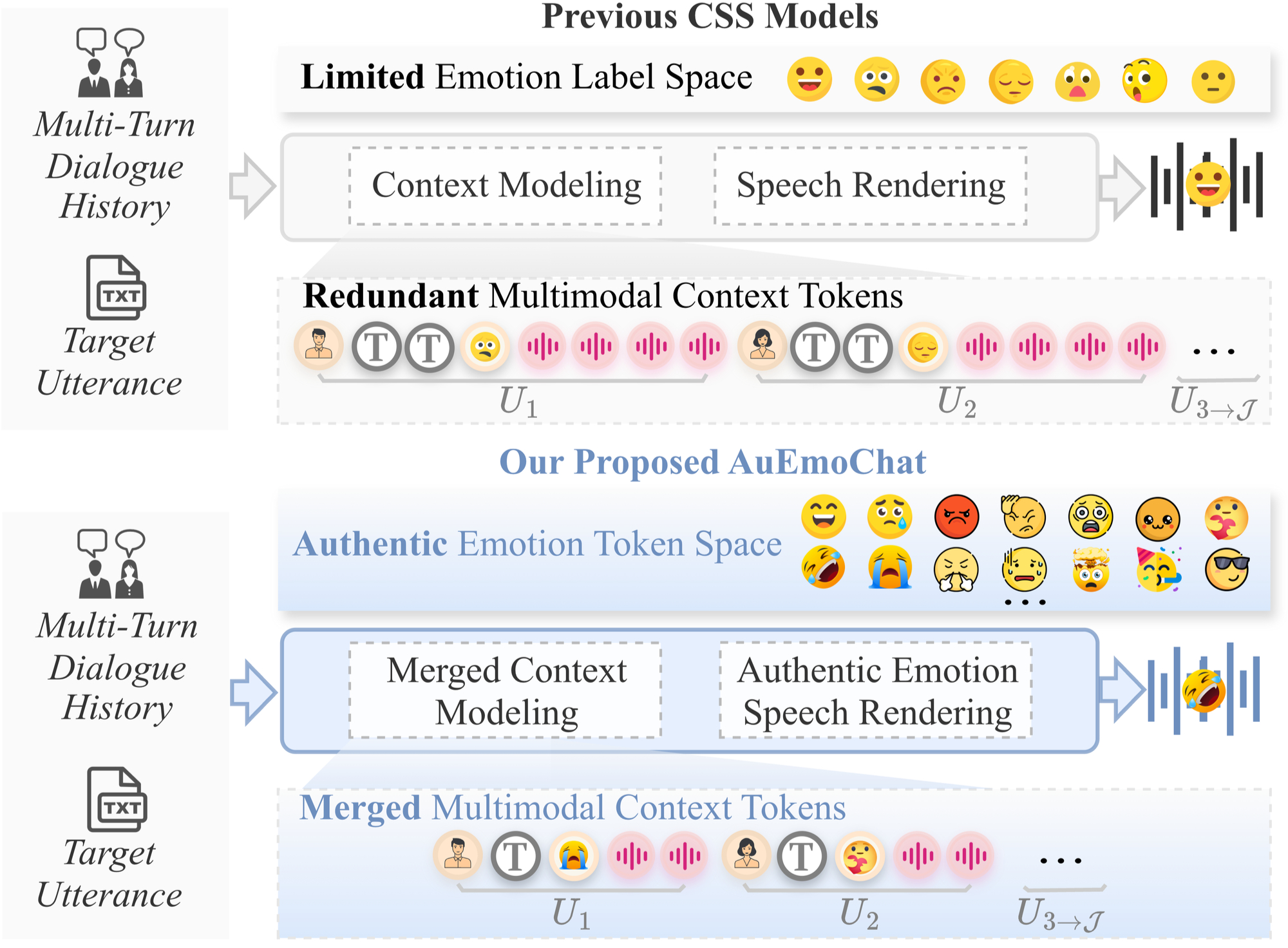}
    \caption{Previous CSS models rely on a limited emotion label space and directly model redundant multimodal context tokens, resulting in limited emotional speech expression. In contrast, AuEmoChat introduces an authentic emotion token space and models the context based on merged tokens, thereby generating more authentic emotional speech.}
    % \vspace{-1em}
    \label{fig:idea}
\end{figure}

Previous CSS studies \cite{guo2021conversational, lee2023dailytalk, xue2023m, li2022enhancing, li2022inferring, deng2024concss, liu2024emphasis, liu2024emotion, jia2025intra, jia2025multimodal, liu2024generative, hu-etal-2025-chain, hu2025unitalker, liu2026emphasis} mainly rely on a set of predefined emotion labels (including anger, disgust, fear, happiness, sadness, surprise, and neutral) derived from Ekman's basic emotion theory \cite{ekman1992argument}, and directly model the multi-turn dialogue history for context modeling. For instance, ECSS \cite{liu2024emotion} encodes multimodal dialogue history into a heterogeneous graph to improve the emotional expressiveness of the synthesized target speech. GPT-Talker \cite{liu2024generative} converts multimodal information \cite{ning2024breaking, fu2026missing, meng2026tri, zou2026tackling, zou2025gea} in the dialogue history into discrete token sequences containing semantic and style information, and utilizes an autoregressive model to predict target acoustic tokens for synthesizing emotional speech. Recently, Chain-Talker \cite{hu-etal-2025-chain} further generates empathetic captions based on these predefined emotion labels, adopting a three-stage framework that includes emotion understanding, semantic understanding, and empathy rendering to enhance the empathetic expression of the target speech.

Despite the progress of these advanced CSS methods in speech quality and expressiveness, they still face the following limitations, as illustrated in Fig. \ref{fig:idea} (top): \textbf{1) Limited Emotion Label Space}: In real-world dialogues, emotional expressions are richer and broader \cite{zhou2022speech, tang2023emomix, liu2025cmgbench}. For example, emotions related to “happiness” include excitement, pleasure, and tearful joy. Such a broad emotional space is more capable of reflecting humans’ authentic emotions (AuEmo). However, conventional predefined emotion labels cannot distinguish these richer and more diverse emotional states, limiting a model's ability to learn authentic emotional expressions. \textbf{2) Redundant Multimodal Context Tokens}: As dialogue turns increase, multimodal token sequences become increasingly long, inevitably introducing a large amount of redundant information \cite{liu2024generative, hu-etal-2025-chain}. These redundant tokens introduce significant noise into contextual modeling, severely interfering with emotion understanding and high-quality speech generation for the target utterance \cite{zhang2025enhancing}.

To address these issues, we propose \textbf{AuEmoChat}, a CSS framework for authentic emotion understanding and rendering. Specifically, we first develop \textit{AuEmoCodec}, which learns a discrete authentic emotion token space from large-scale emotional speech. AuEmoCodec quantizes emotional speech into an AuEmo token in an emotion codebook via Finite Scalar Quantization (FSQ) \cite{mentzer2023finite}. The AuEmo token is then used to reconstruct the perceived emotion scores of the source speech for seven basic emotion axes, enabling the model to capture more authentic and fine-grained emotional states. Second, we design an AuEmo-guided token merging (\textit{AuEmoToMe}) algorithm that merges redundant text and speech tokens in multimodal dialogue history. This design reduces the interference caused by redundant tokens during emotion modeling while preserving emotion-relevant contextual information. As a result, the model can better infer the target AuEmo token and speech tokens for the target utterance. Finally, we propose \textit{Authentic Emotion Flow Matching}, which generates emotionally expressive speech conditioned on the merged dialogue context and the predicted AuEmo token. In addition, this mechanism incorporates AuEmo classifier guidance during the flow matching process, enabling the synthesized speech to better align with the target authentic emotion. In summary, the main contributions of this work are as follows:

\begin{itemize}
    \item[1)] \textit{We propose AuEmoChat}, the first CSS framework dedicated to authentic emotion understanding and rendering.
    \item[2)] \textit{We develop AuEmoCodec}, which leverages finite scalar quantization to learn an authentic emotion token space from large-scale emotional speech, thereby providing more authentic emotional representations.
    \item[3)] \textit{We design AuEmoToMe}, an authentic-emotion-guided token merging algorithm that merges redundant multimodal context tokens while preserving emotion-relevant context. Furthermore, \textit{we propose Authentic Emotion Flow Matching} to improve context-consistent emotional speech rendering.
    \item[4)] Extensive experiments on the NCSSD-EmCap dataset demonstrate that \textit{AuEmoChat significantly outperforms advanced baselines} in both emotional expressiveness and speech quality.
\end{itemize}

\section{Related Works}
\subsection{Speech Neural Codec}
Speech neural codecs convert continuous speech waveforms into discrete tokens, providing the foundation for token-based speech modeling and large language model (LLM)-based audio processing \cite{van2017neural, liu2024semanticodec}. SpeechTokenizer \cite{zhang2023speechtokenizer} uses an encoder-decoder architecture with residual vector quantization (RVQ) to model semantic and acoustic information in speech jointly. SemantiCodec \cite{liu2024semanticodec} adopts a dual-encoder architecture to extract ultra-low-bitrate and semantically rich audio tokens for speech, general sounds, and music. FACodec \cite{ju2024naturalspeech} decomposes speech into content, prosody, timbre, and acoustic-detail subspaces through factorized vector quantization (FVQ), enabling separate generation of different speech attributes. TF-Codec \cite{jiang2023latent} integrates latent-domain predictive coding and learnable time-frequency compression into VQ for low-latency speech coding. CosyVoice \cite{du2024cosyvoice} extracts semantically rich speech tokens with an Automatic Speech Recognition (ASR)-based self-supervised method, while CosyVoice2 \cite{du2024cosyvoice2} further improves token representation by introducing FSQ. Our AuEmoCodec has the following key distinctions: 1) It quantizes emotional speech into fixed AuEmo tokens rather than semantic or acoustic-related tokens. 2) Unlike conventional codecs that rely on ASR supervision to learn semantic information or reconstruct the original speech waveform, AuEmoCodec is trained to reconstruct quantized AuEmo tokens into a richer emotion perception space by modeling perceived scores across seven basic emotion axes, allowing the tokens to capture authentic human-like emotion.

\subsection{Emotion Label Space}
The emotion label space is commonly used to quantify emotional representations in speech and serves as a control condition for emotional speech synthesis \cite{liu2024emotion, hu2025unitalker}. Traditional CSS methods learn emotions based on Ekman’s basic emotion theory \cite{ekman1992argument}, which categorizes emotions into six basic types: anger, disgust, fear, happiness, sadness, and surprise. Emotions that cannot be clearly categorized are usually labeled as neutral. This type of model aligns better with people’s intuitive understanding of emotions in daily life. However, its label space is limited and cannot fully cover the rich and subtle emotional states in real dialogues. Plutchik \cite{plutchik2001nature} points out that humans can express about 34,000 distinct emotions, which further indicates that traditional discrete models with limited categories cannot fully represent the authentic emotion space. To address the limitation of label space, Lian et al. \cite{lian2024open} introduce open-vocabulary emotion recognition, which expands emotion categories through open-vocabulary emotion labels. However, although it increases the number of categories, it also introduces many semantically similar or synonymous labels, failing to effectively disentangle the emotion representation space and thus increasing modeling difficulty. To construct a more authentic emotion token space, we use AuEmoCodec to learn AuEmo tokens from large-scale emotional speech. Specifically: 1) We build a discrete emotion codebook, where each token represents a distinct authentic emotional state, providing a finer-grained representation than traditional emotion categories. 2) Instead of relying on predefined emotion labels, AuEmoCodec learns the AuEmo token representation directly from large-scale emotional speech.

\begin{figure*}
    \centering
    \includegraphics[width=1\linewidth]{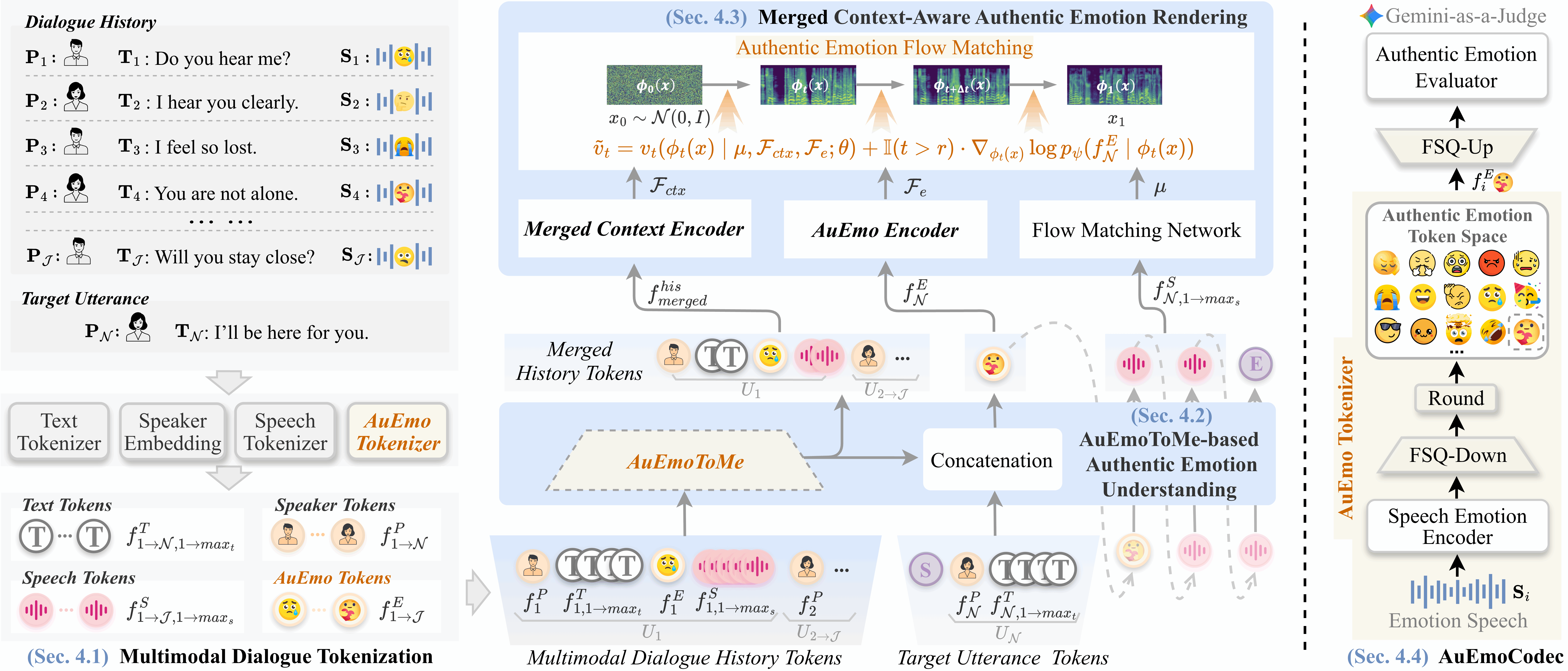}
    \caption{The left side illustrates the overall framework of the proposed AuEmoChat, which includes: Multimodal Dialogue Tokenization, AuEmoToMe-based Authentic Emotion Understanding, and Merged Context-Aware Authentic Emotion Rendering. The right side illustrates the overall framework of the proposed AuEmoCodec.}
    % \vspace{-0.5em}
    \label{fig:auemochat}
\end{figure*}

\subsection{Token Merging}
Token merging improves model efficiency and performance by combining redundant tokens in multimodal sequences \cite{bolya2022token}. For instance, ToMe \cite{bolya2022token} gradually merges similar tokens within the Transformer, enhancing inference speed while largely preserving performance. Token Merger \cite{feng2023efficient} identifies meta tokens representing meaningful image cues and merges similar tokens adaptively, better preserving semantic context and generalizing across vision tasks \cite{zhang2025amns}. FastAST \cite{behera2024fastast} applies token merging to audio spectrograms, improving inference speed while maintaining recognition accuracy. In multimodal dialogue history, speech tokens typically have high bitrate, and long conversations further increase sequence length, making emotional information highly redundant. Redundant tokens dilute model attention to key emotional cues, hindering accurate emotion understanding. Inspired by these works, we propose AuEmoToMe in AuEmoChat: 1) It is the first attempt to introduce token merging into AR-based CSS for modeling multimodal dialogue context. 2) It computes intra-modal similarity among text tokens and speech tokens in the multimodal dialogue history and merges redundant tokens. 3) To preserve emotional information, AuEmoToMe adopts an AuEmo-guided strategy that prioritizes merging redundant tokens while minimally affecting emotion-related key tokens.

\section{Task Definition}
A dialogue can be defined as a sequence of alternating interactions between the user and the agent, denoted as $\{U_1, U_2, \ldots, U_{\mathcal{J}}, U_{\mathcal{N}}\}$, where $\{U_1, U_2, \ldots, U_{\mathcal{J}}\}$ denotes the dialogue history, and $U_{\mathcal{N}}$ denotes the target utterance to be processed by the agent. For each historical utterance $U_i$, it contains multiple modalities $\{\text{P}_i, \text{T}_i, \text{S}_i\}$, which represent the speaker, text, and speech of the $i$-th turn, respectively. For the target utterance $U_{\mathcal{N}}$, the speaker $\text{P}_{\mathcal{N}}$ and text $\text{T}_{\mathcal{N}}$ are given, and the goal is to generate the corresponding speech $\text{S}_{\mathcal{N}}$. CSS based on authentic emotion understanding and rendering needs to address the following points: \textbf{1)} How to more accurately represent the authentic emotion of each utterance in the dialogue. \textbf{2)} How to effectively infer the authentic emotion token and speech token of the target utterance from a long multimodal dialogue history. \textbf{3)} How to synthesize speech with authentic emotion expression that is consistent with the current dialogue context.

\section{Methodology: AuEmoChat}
Fig. \ref{fig:auemochat} (left) illustrates the overall architecture of \textbf{AuEmoChat}. The \textbf{Multimodal Dialogue Tokenization} module extracts text, speech, speaker, and AuEmo tokens from the input dialogue. The \textbf{AuEmoToMe-based Authentic Emotion Understanding} module incorporates the proposed AuEmoToMe into the LLM to merge redundant text and speech tokens in the dialogue history, based on which the LLM predicts the AuEmo token and speech tokens of the target utterance. The \textbf{Merged Context-Aware Authentic Emotion Rendering} module conditions on the merged history tokens, the predicted AuEmo token, and the speech tokens to synthesize authentic emotional speech that is consistent with the context through an authentic emotion flow matching mechanism. In addition, Fig. \ref{fig:auemochat} (right) illustrates the training process of \textit{AuEmoCodec}, which will be described in detail at the end of this section.

\subsection{Multimodal Dialogue Tokenization}
In this section, we introduce how multimodal dialogue data are processed to extract the corresponding multimodal tokens. The implementation details are as follows.

\textbf{Text Tokenizer.} We  adopt a BPE-based Text Tokenizer \cite{gage1994new} to tokenize the dialogue text.
% This design avoids an additional grapheme-to-phoneme (G2P) transformation step and allows the model to learn text representations directly from the context. 
Specifically, as shown in Eq. \ref{eq:1},  the input dialogue text sequence ${\text{T}_{1 \rightarrow \mathcal{N}}}$ is processed by the Text Tokenizer to obtain the corresponding text token embeddings:
\begin{equation}
    f^{T}_{1 \rightarrow \mathcal{N}, 1 \rightarrow max_{t}} = \text{Text Tokenizer}(\text{T}_{1 \rightarrow \mathcal{N}})
    \label{eq:1}
\end{equation}
where ``$1 \rightarrow \mathcal{N}$'' denotes the dialogue turn indices, and ``$1 \rightarrow max_{t}$'' denotes the text token indices within each utterance.

\textbf{Speaker Embedding.} 
We use Speaker Embedding to encode the discrete identities of different speakers. Specifically, the speaker sequence ${\text{P}_{1 \rightarrow \mathcal{N}}}$ of the current dialogue is processed by the Speaker Embedding to obtain the corresponding speaker token embeddings:
\begin{equation}
    f^{P}_{1 \rightarrow \mathcal{N}} = \text{Speaker Embedding}(\text{P}_{1 \rightarrow \mathcal{N}})
    \label{eq:2}
\end{equation}

\textbf{Speech Tokenizer.} We adopt the Speech Tokenizer proposed in CosyVoice2 \cite{du2024cosyvoice2} to discretize the speech signals in the dialogue history. 
% This tokenizer integrates the FSQ module into the SenseVoice-Large ASR encoder \cite{an2024funaudiollm}, so that the intermediate speech representations produced by the encoder can be quantized into discrete speech tokens and further mapped into speech token embeddings. 
Specifically, the dialogue history speech sequence ${\text{S}_{1 \rightarrow \mathcal{J}}}$ is processed by the Speech Tokenizer to obtain the corresponding speech token embeddings:
\begin{equation}
    f^{S}_{1 \rightarrow \mathcal{J}, 1 \rightarrow max_{s}} = \text{Speech Tokenizer}(\text{S}_{1 \rightarrow \mathcal{J}})
    \label{eq:3}
\end{equation}
where ``$1 \rightarrow max_{s}$'' denotes the speech token indices within each utterance.

\textcolor[rgb]{0.8,0.4,0.0}{\textbf{AuEmo Tokenizer.}} To represent authentic emotional expressions in dialogue, we obtain an AuEmo Tokenizer by training AuEmoCodec, as shown in Fig. \ref{fig:auemochat} (right). AuEmoCodec is trained on large-scale emotional speech data to learn a discrete authentic emotion token space via FSQ \cite{ren2025amdo}. We construct an emotion codebook of size 1000, from which 750 tokens are activated after training and used as AuEmo tokens. The detailed training procedure of \textit{AuEmoCodec} is described in Sec. \ref{auemocodectrain}. The dialogue history speech sequence ${\text{S}_{1 \rightarrow \mathcal{J}}}$ is then processed by the AuEmo Tokenizer to obtain the corresponding AuEmo token embeddings:
\begin{equation}
    f^{E}_{1 \rightarrow \mathcal{J}} = \text{AuEmo Tokenizer}(\text{S}_{1 \rightarrow \mathcal{J}})
    \label{eq:4}
\end{equation}

\subsection{AuEmoToMe-based Authentic Emotion Understanding}
In AR-based CSS, multimodal dialogue history is typically organized as a unified token sequence for contextual modeling. As dialogue turns increase, text and speech tokens accumulate rapidly, resulting in a long multimodal input sequence with many redundant tokens. To address this issue, we design the AuEmoToMe-based Authentic Emotion Understanding module, which merges redundant emotion-related text and speech tokens in the dialogue history and predicts the target AuEmo and speech tokens from the merged history representation. The module consists of three execution steps:

\textbf{Multimodal Dialogue Token Sequence Construction.} Following the order of dialogue turns, the dialogue history is constructed as the following multimodal dialogue history token sequence:
\begin{equation}
\begin{aligned}
U_{1 \rightarrow \mathcal{J}} =
\{ & f^P_1, f^T_{1,1 \rightarrow max_t}, f^E_1, f^S_{1,1 \rightarrow max_s}, \\
   & \dots, \\
   & f^P_{\mathcal{J}}, f^T_{\mathcal{J},1 \rightarrow max_t}, f^E_{\mathcal{J}}, f^S_{\mathcal{J},1 \rightarrow max_s} \}
\end{aligned}
\label{eq:history_seq}
\end{equation}
where $f^P_i$, $f^T_{i,1 \rightarrow max_t}$, $f^E_i$, and $f^S_{i,1 \rightarrow max_s}$ denote the speaker token, text tokens, AuEmo token, and speech tokens of the $i$-th utterance.

For the target utterance, we only use the speaker token and text tokens as conditional inputs to construct the target utterance token sequence:
\begin{equation}
U_{\mathcal{N}} = \{\tikz[baseline=(char.base)]{
  \node[draw,circle,inner sep=1pt] (char) {S};
}, f^P_{\mathcal{N}}, f^T_{\mathcal{N},1 \rightarrow max_t}\}
\label{eq:target_seq}
\end{equation}
where $\tikz[baseline=(char.base)]{
  \node[draw,circle,inner sep=1pt] (char) {S};
}$ denotes the start token of the target utterance.

\textbf{Authentic Emotion-driven Token Merging (\textcolor[rgb]{0.8,0.4,0.0}{AuEmoToMe}).} 
To reduce the interference of redundant information in multi-turn multimodal dialogue history on target emotion modeling, we design AuEmoToMe based on ToMe \cite{bolya2022token}. It computes cosine similarity between tokens to identify redundant text and speech tokens, and merges them into a more compact merged history tokens sequence. To preserve authentic emotional information during compression, AuEmoToMe employs an AuEmo-guided merging strategy, where the AuEmo token of each historical utterance acts as an emotion anchor, guiding the weighted aggregation of matched text and speech redundant tokens separately. Thus, AuEmoToMe preserves more emotion-relevant information while merging redundant tokens.

% \begin{algorithm}[t]
% \caption{AuEmoToMe}
% \label{alg:auemotome}
% \KwIn{Dialogue history tokens $U_{1 \rightarrow \mathcal{J}}$, merge ratio $\rho$}
% \KwOut{Merged history tokens $f^{his}_{merged}$}

% Initialize $f^{his}_{merged} \leftarrow \emptyset$\;

% \For{each utterance $U_i$ in $U_{1 \rightarrow \mathcal{J}}$}{
%     1. Extract text tokens $f^T_{i,1\rightarrow max_t}$, speech tokens $f^S_{i,1\rightarrow max_s}$, and AuEmo token $f^E_i$.
    
%     2. Construct token sequence $X^M_i$, where $M \in \{T, S\}$ denotes text or speech.
    
%     3. Partition $X^M_i$ into source set $A^M$ and destination set $B^M$, where $|B^M|=\lfloor |X^M_i|(1-\rho) \rfloor$ and $A^M = X^M_i \setminus B^M$.

%     4. Compute cosine similarity between tokens in $A^M$ and tokens in $B^M$.
    
%     5. Match each token in $A^M$ to its similar token in $B^M$.
    
%     6. Select the top-$k$ pairs with the highest similarity.
    
%     \For{each selected pair $(a_u, b_v) \in (A^M, B^M)$}{
%         Compute cosine similarities $s_a=\mathrm{sim}(a_u,f^E_i)$ and $s_b=\mathrm{sim}(b_v,f^E_i)$.
        
%         Normalize them into fusion weights:
%         $w_a=\frac{\exp(s_a)}{\exp(s_a)+\exp(s_b)}$, 
%         $w_b=\frac{\exp(s_b)}{\exp(s_a)+\exp(s_b)}$.
        
%         Update token representation: $b_v \leftarrow w_a a_u + w_b b_v$.
%     }
    
%     Update $U_i$ with the merged tokens $X^M_i$.

%     Append the updated utterance $U_i$ to $f^{his}_{merged}$.
% }

% \end{algorithm}

Specifically, AuEmoToMe takes the dialogue history token sequence $U_{1 \rightarrow \mathcal{J}}$ as input and performs token merging for each historical utterance. For each utterance $U_i$, its text tokens $f^T_{i,1 \rightarrow max_t}$ and speech tokens $f^S_{i,1 \rightarrow max_s}$ are separately used to construct token sequences $X^M_i$, where $M \in \{T, S\}$ denotes text or speech. The AuEmo token $f^E_i$ serves as an emotion anchor. Given the merging ratio $\rho$, $X^M_i$ is partitioned into a source set $A^M$ and a destination set $B^M$. Cosine similarities between tokens in $A^M$ and $B^M$ are computed to select the top-$k$ matched pairs for merging. The selected pairs are then weighted by their similarities to the AuEmo anchor and aggregated accordingly. Finally, the merged tokens are restored to their original temporal order, and the merged tokens of all utterances are concatenated to form the merged dialogue history tokens $f^{his}_{merged}$.

\textbf{Target AuEmo Token and Speech Token Inference.}
The merged history tokens are concatenated with the target utterance token sequence:
\begin{equation}
\{ f^{his}_{merged},
\tikz[baseline=(char.base)]{\node[draw,circle,inner sep=1pt] (char) {S};},
f^P_{\mathcal{N}},
f^T_{\mathcal{N},1 \rightarrow max_t} \}
\label{eq:llm_input}
\end{equation}

The model first predicts the AuEmo token of the target utterance $f^E_{\mathcal{N}}$, and then generates the speech token sequence $f^S_{\mathcal{N},1 \rightarrow max_s}$, which is terminated by an end token $\tikz[baseline=(char.base)]{
  \node[draw,circle,inner sep=1pt] (char) {E}}$. Therefore, the target output sequence can be defined as:
\begin{equation}
\mathcal{Y}_{\mathcal{N}} =
\{f^E_{\mathcal{N}}, f^S_{\mathcal{N},1 \rightarrow max_s}, \tikz[baseline=(char.base)]{
  \node[draw,circle,inner sep=1pt] (char) {E}}\}
\label{eq:target_output}
\end{equation}

The generation process follows a conditional autoregressive formulation:
\begin{equation}
\begin{aligned}
p(\mathcal{Y}_{\mathcal{N}} \mid f^{his}_{merged}, \tikz[baseline=(char.base)]{
  \node[draw,circle,inner sep=1pt] (char) {S}}, f^P_{\mathcal{N}}, 
f^T_{\mathcal{N},1 \rightarrow max_t}; \Theta) \
\end{aligned}
\label{eq:ar_factorization}
\end{equation}
where $\Theta$ denotes the model parameters.

To explicitly model the authentic emotion understanding process, the prediction of the target AuEmo token is formulated as:
\begin{equation}
p(f^E_{\mathcal{N}} \mid f^{his}_{merged}, \tikz[baseline=(char.base)]{
  \node[draw,circle,inner sep=1pt] (char) {S}}, f^P_{\mathcal{N}}, f^T_{\mathcal{N},1 \rightarrow max_t}; \Theta)
\label{eq:emo_pred}
\end{equation}
and the speech token sequence is then generated conditioned on the predicted emotion token:
\begin{equation}
p(f^S_{\mathcal{N},1 \rightarrow max_s}
\mid
f^{his}_{merged}, \tikz[baseline=(char.base)]{
  \node[draw,circle,inner sep=1pt] (char) {S}}, f^P_{\mathcal{N}}, f^T_{\mathcal{N},1 \rightarrow max_t}, f^E_{\mathcal{N}}; \Theta)
\label{eq:speech_pred}
\end{equation}

\subsection{Merged Context-Aware Authentic Emotion Rendering}
This module aims to generate conversational speech with authentic emotional expression by jointly conditioning on the merged history tokens $f^{his}_{merged}$, the target AuEmo token $f^{E}_{\mathcal{N}}$, and the target speech token sequence $f^{S}_{\mathcal{N},1 \rightarrow max_s}$.

\textbf{Merged Context Encoder.} To obtain a context condition for Flow Matching, we feed $f^{his}_{merged}$ into the Merged Context Encoder. This encoder uses a bidirectional GRU \cite{chung2014empirical} to model the merged history tokens and capture the global dialogue context. The resulting context condition can be formulated as:
\begin{equation}
\mathcal{F}_{ctx} = \text{Merged Context Encoder}(f^{his}_{merged})
\label{eq:ctx_encoder}
\end{equation}

\textbf{AuEmo Encoder.} To obtain an authentic emotion condition for Flow Matching, we feed the predicted AuEmo token $f^{E}_{\mathcal{N}}$ into the AuEmo Encoder. This encoder projects the AuEmo token into the conditioning space of the Flow Matching model. The resulting authentic emotion condition can be formulated as:
\begin{equation}
\mathcal{F}_{e} = \text{AuEmo Encoder}(f^{E}_{\mathcal{N}})
\label{eq:auemo_encoder}
\end{equation}

\textbf{\textcolor[rgb]{0.8,0.4,0.0}{Authentic Emotion Flow Matching.}} 
Given the predicted target speech token sequence $f^{S}_{\mathcal{N},1 \rightarrow max_s}$, we first extract its acoustic prior representation, denoted as $\mu$. The context condition $\mathcal{F}_{ctx}$ and the target emotion condition $\mathcal{F}_e$ are then injected into the Flow Matching network together with $\mu$, such that the generation process is jointly constrained by the target speech prior, the merged dialogue context, and the target authentic emotion.

Following conditional Flow Matching, let $x_1$ denote the ground-truth mel-spectrogram of the target utterance. Let $x_0 \sim \mathcal{N}(0, I)$ denote the Gaussian noise. We define the interpolation path between $x_0$ and $x_1$ as:
\begin{equation}
\phi_t(x) = \big(1 - (1-\sigma_{min})t\big)x_0 + t x_1, \quad t \in [0,1]
\label{eq:fm_path}
\end{equation}
where $\sigma_{\min}=10^{-6}$, following CosyVoice2 \cite{du2024cosyvoice2}. The corresponding target vector field is:
\begin{equation}
u_t(\phi_t(x)\mid x_1) = x_1 - (1-\sigma_{min})x_0
\label{eq:target_vector_field}
\end{equation}

Conditioned on $\mu$, $\mathcal{F}_{ctx}$, and $\mathcal{F}_e$, the Flow Matching network predicts the vector field:
\begin{equation}
v_t(\phi_t(x) \mid \mu, \mathcal{F}_{ctx}, \mathcal{F}_e; \theta)
\label{eq:base_vector_field}
\end{equation}
where $\theta$ denotes the model parameters. The main objective is to make the predicted vector field match the target transport direction from noise to the real mel-spectrogram. Therefore, the Flow Matching loss ($\mathcal{L}_{fm}$) is defined as:
\begin{equation}
\mathcal{L}_{fm}
=
\mathbb{E}_{x_0,x_1,t}
\left[
\left\|
v_t(\phi_t(x) \mid \mu, \mathcal{F}_{ctx}, \mathcal{F}_e; \theta)
-
u_t(\phi_t(x)\mid x_1)
\right\|_2^2
\right]
\label{eq:fm_loss}
\end{equation}

Although the $\mathcal{L}_{fm}$ learns the acoustic transport path, it does not explicitly constrain the intermediate mel states to preserve the target authentic emotion. To address this issue, we further introduce an AuEmo classifier guidance mechanism. Specifically, an auxiliary AuEmo classifier $\psi$ is applied to the intermediate state $\phi_t(x)$ to estimate the probability that the current mel state matches the target AuEmo token $f^{E}_{\mathcal{N}}$. Based on this classifier, the guided vector field can be written as:
\begin{equation}
\tilde{v}_t
=
v_t(\phi_t(x) \mid \mu, \mathcal{F}_{ctx}, \mathcal{F}_e; \theta)
+
\mathbb{I}(t > r)\cdot
\nabla_{\phi_t(x)}
\log p_{\psi}\big(f^{E}_{\mathcal{N}} \mid \phi_t(x)\big)
\label{eq:guided_vector_field}
\end{equation}
where $\mathbb{I}(\cdot)$ is the indicator function, $r$ is a activation threshold set to $0.7$, and $\psi$ denotes the pre-trained AuEmoCodec.

The purpose is to guide the intermediate mel state toward a direction that is more consistent with the target authentic emotion. In this way, the optimization trajectory is guided not only by the acoustic transport objective, but also by the target authentic emotion. Notably, we activate this guidance only when $t > r$, because the intermediate state is still close to Gaussian noise when $t$ is small and does not yet contain stable emotion-related spectro-temporal patterns. Applying emotion guidance too early may introduce noisy gradients and weaken the learning of the basic transport path. When $t > r$, the intermediate mel state becomes more structured, and the emotion classifier can provide more reliable guidance.

During inference, the model starts from Gaussian noise $x_0 \sim \mathcal{N}(0, I)$ and solves the following ordinary differential equation:
\begin{equation}
\frac{d\phi_t(x)}{dt}
=
v_t(\phi_t(x) \mid \mu, \mathcal{F}_{ctx}, \mathcal{F}_e; \theta), \quad t \in [0,1]
\label{eq:ode_sampling}
\end{equation}
The final output $x_1$ is the generated mel-spectrogram, which is then converted into a waveform by the HiFi-GAN vocoder \cite{kong2020hifi}.

\subsection{Training Process of AuEmoCodec}
\label{auemocodectrain}
AuEmoCodec aims to learn discrete AuEmo tokens from large-scale emotional speech data that are more discriminative and more consistent with real human emotional expression. Specifically, given an emotional speech segment $S_i$, we first use a Speech Emotion Encoder to extract its emotion representation. We then map this representation into a low-rank quantization space through FSQ-Down and discretize it with a bounded rounding operation. In our implementation, the quantization levels are set to $\text{levels}=[8,5,5,5]$. The resulting discrete code is defined as the AuEmo token:
\begin{equation}
f^{E}_{i} = \mathrm{ROUND}(\mathrm{FSQ\mbox{-}Down}(\text{Speech Emotion Encoder}(S_i)))
\label{eq:auemo_token}
\end{equation}

To enable AuEmoCodec to focus exclusively on emotion representation learning, we do not train it to reconstruct the original waveform. Instead, motivated by the fact that authentic human emotion is often expressed as a complex mixture of multiple affective states \cite{zhou2022speech, tang2023emomix}, we use the perceived scores of speech sample $S_i$ over seven basic emotion axes as the reconstruction target. Specifically, the quantized AuEmo token $f^E_i$ is first mapped back to a high-dimensional representation through FSQ-Up, from which the model reconstructs the multi-axis perceived emotion scores, denoted as $P^{ae}_i$. To obtain supervision targets, we use Gemini-2.5-Flash \cite{comanici2025gemini} to annotate the perceived scores of input speech along each emotion axis, denoted as $GT_i^{ae}$. The prompt used for Gemini-2.5-Flash, together with representative examples of the resulting perceived scores over the seven emotion axes, is provided in Appendix A. AuEmoCodec is then trained by minimizing the mean squared error (MSE) between $P_i^{ae}$ and $GT_i^{ae}$, which encourages the learned AuEmo tokens to capture emotional nuances and thereby establish a more effective authentic emotion token space for CSS.
% To make AuEmoCodec to focus on emotion representation learning, we do not train it to reconstruct the original waveform. Instead, motivated by the fact that authentic human emotion is often expressed as a combination or mixture of multiple basic emotions \cite{zhou2022speech, tang2023emomix}, we use the perceived scores of the input speech over seven basic emotion axes as the reconstruction target. Specifically, the quantized AuEmo token $f^E_i$ is first mapped back to a high-dimensional representation through FSQ-Up, from which the model predicts the perceived scores of seven basic emotions, denoted as $P^{ae}_i$. To obtain supervision targets, we use Gemini-2.5-Flash \cite{comanici2025gemini} to annotate the perceived emotion scores of real speech, and denote its output as $GT^{ae}_i$. The prompt used for Gemini-2.5-Flash, together with representative examples of the resulting perceived scores over the seven basic emotions, is provided in Appendix A. AuEmoCodec is then trained by minimizing the mean squared error (MSE) between $P_i^{ae}$ and $GT_i^{ae}$, encouraging the learned AuEmo tokens to capture more authentic and fine-grained emotional states and thus forming a more effective authentic emotion token space for CSS.

\section{Experiments}
This section presents the experimental setup, baseline and ablation models, and the evaluation metrics used to assess speech quality and emotional expressiveness. Additional experimental details are provided in Appendix B.

\subsection{Experimental Setup}
We use the open-source NCSSD-EmCap \cite{hu-etal-2025-chain} dataset to evaluate the effectiveness of AuEmoChat. This dataset integrates three high-quality dialogue speech datasets: DailyTalk \cite{lee2023dailytalk}, NCSSD \cite{liu2024generative}, and MultiDialog \cite{park2024let}. Specifically, NCSSD-EmCap contains approximately 384 hours of speech data. It comprises 18,580 dialogues with a total of 245,984 utterances, averaging 13.2 turns per dialogue, and covers 25 speakers. Among them, 124,599 utterances are from male speakers and 121,385 utterances are from female speakers. The dataset is divided into training, validation, and test sets with a ratio of 8:1:1, and the same split is used for training both AuEmoCodec and AuEmoChat. For training, AuEmoChat uses 4 NVIDIA A100 GPUs with a batch size of 4 and 8 gradient accumulation steps.

\begin{table*}[t]
  \centering
  \caption{Subjective (95\% confidence interval \cite{yasuda23_interspeech}) and objective results with different comparative models. \colorbox{blue_1}{Blue} indicates the best performance.}
  \label{tab:baselines}
  \small
  \resizebox{1\textwidth}{!}{
  \begin{tabular}{l|ccccccc} 
    \toprule
    \textbf{Methods} & \textbf{N-DMOS ($\uparrow$)} & \textbf{E-DMOS ($\uparrow$)} & \textbf{WER ($\downarrow$)} & \textbf{MCD ($\downarrow$)} & \textbf{SpkSIM ($\uparrow$)} & \textbf{EmoACC ($\uparrow$)} & \textbf{AuEmoACC ($\uparrow$)}\\
    \midrule

    BaseCSS \cite{lee2023dailytalk} \textcolor{gray!77}{(\textit{ICASSP 2023})}  & 3.431 $\pm$ 0.023 & 3.299 $\pm$ 0.024 & 27.28 & 9.359 & 70.75 & 45.62 & 19.12\\
    ECSS \cite{liu2024emotion} \textcolor{gray!77}{(\textit{AAAI 2024})} & 3.675 $\pm$ 0.022 & 3.509 $\pm$ 0.022 & 25.10 & 8.568 & 72.51 & 49.85 & 22.31    \\
    GPT-Talker \cite{liu2024generative}  \textcolor{gray!77}{(\textit{MM 2024})} & 3.812 $\pm$ 0.023 & 3.654 $\pm$ 0.022 & 20.49 & 8.061 & 77.68 & 57.08 & 23.45 \\ 
    Chain-Talker \cite{hu-etal-2025-chain} \textcolor{gray!77}{(\textit{ACL 2025})} & 3.955 $\pm$ 0.025 & 3.756 $\pm$ 0.018 & 15.07 & 7.686 & 77.56 & 57.16 & 24.12 \\
    \midrule
    \textbf{AuEmoChat (Ours)}              & \cellcolor{blue_1}\textbf{4.171 $\pm$ 0.026} & \cellcolor{blue_1}\textbf{3.979 $\pm$ 0.021} & \cellcolor{blue_1}\textbf{9.14}  & \cellcolor{blue_1}\textbf{6.847}  & \cellcolor{blue_1}\textbf{78.03} & \cellcolor{blue_1}\textbf{61.04} & \cellcolor{blue_1}\textbf{28.71} \\
    \bottomrule
  \end{tabular}
  }
\end{table*}

\subsection{Baseline and Ablation Models}
\textbf{Baseline Models.} To evaluate the effectiveness of AuEmoChat, we compare it with state-of-the-art (SOTA) emotional CSS systems, including \textbf{1) \textit{BaseCSS}} \cite{guo2021conversational, lee2023dailytalk}, \textbf{2) \textit{ECSS}} \cite{liu2024emotion}, \textbf{3) \textit{GPT-Talker}} \cite{liu2024generative}, and \textbf{4) \textit{Chain-Talker}} \cite{hu-etal-2025-chain}. For fair comparison, all baseline models are configured to use the authentic emotion token space.

\textbf{Ablation Models.} To assess the contribution of each component in AuEmoChat, we conduct three groups of ablation studies. \textbf{1) Ablations on AuEmo Tokenizer:} We replace AuEmo Tokenizer with three alternative emotion representations to examine the role of the authentic emotion token space in supporting emotion understanding and rendering: \textbf{\textit{Abl.1 w/ LimEmo}}, which uses a traditional limited emotion label space. \textbf{\textit{Abl.2 w/ OVEmo}}, which uses an open-vocabulary (OV) emotion label space. \textbf{\textit{Abl.3 w/ EmoCap}}, which uses emotion captions. \textbf{2) Ablations on AuEmoToMe:} We remove AuEmoToMe or its AuEmo-guided merging strategy to investigate its importance for target-utterance emotion understanding and speech token prediction: \textbf{\textit{Abl.4 w/o AuEmoToMe}} removes the entire AuEmoToMe module. \textbf{\textit{Abl.5 w/o AuEmo-guided Strategy}} removes the AuEmo-guided token merging strategy. \textbf{3) Ablations on Authentic Emotion Flow Matching:} We remove different conditioning components to analyze the roles of merged dialogue context and authentic emotion constraints in context-consistent emotional speech rendering: \textbf{\textit{Abl.6 w/o FM-AuEmo}} removes the AuEmo condition. \textbf{\textit{Abl.7 w/o FM-ACG}} removes AuEmo classifier guidance (ACG). \textbf{\textit{Abl.8 w/o FM-Context}} removes merged context condition.
% Detailed descriptions of the baseline models and ablation settings are provided in Appendices B.2 and B.3.

\subsection{Evaluation Metrics}
We use the following metrics to evaluate each model’s performance:

\textbf{Subjective Metrics:}  Naturalness-DMOS (\textbf{N-DMOS}) \cite{streijl2016mean, liu2024emphasis, liu2025retrieval} assesses the naturalness and overall quality of synthesized speech. Emotion-DMOS (\textbf{E-DMOS}) \cite{liu2024emotion, hu-etal-2025-chain} measures the emotional expressiveness and consistency with ground truth. A total of 30 trained, English-proficient evaluators participated in the subjective evaluation. Each evaluator reviewed the dialogue context and listened to each synthesized sample at least three times before rating its naturalness and emotional consistency.

\textbf{Objective Metrics:} We employ Word Error Rate (\textbf{WER}) \cite{morris2004and}, Mel Cepstral Distortion (\textbf{MCD}) \cite{kubichek1993mel, chen2022v2c}, and Speaker Similarity (\textbf{SpkSIM}) \cite{cong2025emodubber, zhang2024speaker} to evaluate the quality of the synthesized speech. For emotion expressiveness, Emotion Accuracy (\textbf{EmoACC}) evaluates performance under a seven-category basic emotion space using emotion2vec \cite{ma2024emotion2vec}, while Authentic Emotion Accuracy (\textbf{AuEmoACC}) measures consistency in the authentic emotion token space using the trained AuEmoCodec.

\textbf{AuEmoCodec Analysis Metrics:} To validate the training strategy and hyperparameter design of AuEmoCodec, we employ the following metrics: 1) Tolerance at $m$ axes (\textbf{Tol@$m$}) evaluates the reconstruction accuracy of multi-axis perceived emotion scores while allowing at most $m$ mismatched axes. 2) Strict Emotion Axis Alignment (\textbf{S-EAA}) measures the reconstruction accuracy of perceived emotion scores across all emotion axes. 3) Dominant Emotion Axis Alignment (\textbf{D-EAA}) evaluates the accuracy of the reconstructed dominant emotion axis. 4) \textbf{Usage} measures codebook utilization.
% Full details of the subjective, objective, and analytical evaluation metrics are provided in Appendix B.4.

\section{Results and Discussion}
In this section, we comprehensively evaluate AuEmoChat through baseline comparisons, component ablations, analysis of the AuEmoCodec architecture, and analysis of the context token merging rate. Additional experimental results are provided in Appendix C.

\begin{table*}[t]
  \centering
  \caption{Subjective (95\% confidence interval) and objective results with different ablation models. \colorbox{blue_1}{Blue} indicates the best performance, and \underline{underlined} indicates the worst performance.}
  \label{tab:abl}
  \small
  \resizebox{1\textwidth}{!}{
  \begin{tabular}{l|ccccccc} 
    \toprule
    \textbf{Methods} & \textbf{N-DMOS ($\uparrow$)} & \textbf{E-DMOS ($\uparrow$)} & \textbf{WER ($\downarrow$)} & \textbf{MCD ($\downarrow$)} & \textbf{SpkSIM ($\uparrow$)} & \textbf{EmoACC ($\uparrow$)} & \textbf{AuEmoACC ($\uparrow$)}\\
    \midrule
    \rowcolor{gary_1}
    \multicolumn{8}{l}{\textit{\textbf{Ablations on AuEmo Tokenizer}}} \\
    \arrayrulecolor{gray!50} \hline

    \hspace{2mm} \textit{Abl.1 w/ LimEmo} & 3.983 $\pm$ 0.020 & \underline{3.736 $\pm$ 0.025} & 9.69 & 6.921 & 77.59 & 58.46 & \underline{20.38} \\
    \hspace{2mm} \textit{Abl.2 w/ OVEmo} & 3.889 $\pm$ 0.024 & 3.859 $\pm$ 0.022 & 9.74 &  6.971 & 77.46 & 59.19 & 24.57\\
    \hspace{2mm} \textit{Abl.3 w/ EmoCap} & 3.979 $\pm$ 0.022 & 3.765 $\pm$ 0.021 & 9.43 & 6.936 & 77.46 & 58.54 & 24.15\\
    \midrule    
    \rowcolor{gary_1}
    \multicolumn{8}{l}{\textit{\textbf{Ablations on AuEmoToMe}}} \\
    \arrayrulecolor{gray!50} \hline 
    \hspace{2mm} \textit{Abl.4 w/o AuEmoToMe} & \underline{3.817 $\pm$ 0.027} & 3.750 $\pm$ 0.024 & 9.55 & 7.274 & 77.84 & \underline{57.97} & 25.72\\
    \hspace{2mm} \textit{Abl.5 w/o AuEmo-guided Strategy} & 3.948 $\pm$ 0.022 & 3.815 $\pm$ 0.027 & 9.37 & 6.905 & 77.53 & 59.81 & 27.02\\
    \midrule
    \rowcolor{gary_1}
    \multicolumn{8}{l}{\textit{\textbf{Ablations on Authentic Emotion Flow Matching}}} \\
    \arrayrulecolor{gray!50} \hline 
    \hspace{2mm} \textit{Abl.6 w/o FM-AuEmo} & 3.982 $\pm$ 0.023 & 3.774 $\pm$ 0.024 & 9.15 & 6.865 & 77.18 & 58.33 & 23.88\\
    \hspace{2mm} \textit{Abl.7 w/o FM-ACG} & 4.008 $\pm$ 0.023 & 3.820 $\pm$ 0.022 & 9.27 & 6.989 & 77.71 & 59.84 & 25.47 \\
    \hspace{2mm} \textit{Abl.8 w/o FM-Context} & 3.863 $\pm$ 0.024 & 3.873 $\pm$ 0.019 & \underline{13.63} & \underline{7.610} & \underline{72.61} & 60.17 & 26.31\\
    \midrule
    \textbf{AuEmoChat (Ours)}              & \cellcolor{blue_1}\textbf{4.171 $\pm$ 0.026} & \cellcolor{blue_1}\textbf{3.979 $\pm$ 0.021} &  \cellcolor{blue_1}\textbf{9.14}  & \cellcolor{blue_1}\textbf{6.847}  & \cellcolor{blue_1}\textbf{78.03} & \cellcolor{blue_1}\textbf{61.04} & \cellcolor{blue_1}\textbf{28.71} \\
    \bottomrule
  \end{tabular}
  }
\end{table*}

\subsection{AuEmoChat \textit{vs.} Baseline Models}
Table \ref{tab:baselines} presents the subjective and objective evaluation results comparing AuEmoChat with SOTA CSS baselines. Overall, AuEmoChat achieves the best performance across all metrics, demonstrating its superiority in both speech quality and authentic emotion modeling. \textbf{\textit{For subjective evaluations}}, AuEmoChat achieves the highest scores in both N-DMOS (4.171) and E-DMOS (3.979), surpassing the strongest baseline Chain-Talker by 0.216 and 0.223, respectively. This indicates that AuEmoChat generates more natural speech with more accurate and context-consistent emotional expression. \textbf{\textit{For objective evaluations}}, AuEmoChat consistently outperforms all baselines in both speech quality and emotion expressiveness. Compared with the strongest baseline Chain-Talker, AuEmoChat reduces WER from 15.07 to 9.14 and MCD from 7.686 to 6.847, indicating improved pronunciation accuracy and acoustic fidelity. It also achieves the highest speaker similarity (78.03), showing better preservation of speaker identity. In terms of emotional expression, AuEmoChat improves EmoACC from 57.16 to 61.04 and AuEmoACC from 24.12 to 28.71. The improvement in AuEmoACC further demonstrates the effectiveness of the proposed AuEmoChat in capturing authentic emotional states beyond basic emotion labels. Overall, these results confirm that AuEmoChat effectively addresses the limitations of existing CSS systems by improving authentic emotion understanding, reducing redundant context interference, and generating more expressive and context-consistent speech.

\subsection{Ablation Results of Key Components}
Table \ref{tab:abl} presents the ablation results obtained by removing or replacing key components in AuEmoChat. Overall, all ablation variants perform worse than the full AuEmoChat model across both speech quality and emotion expressiveness metrics, confirming the necessity of each proposed component. \textbf{\textit{In Abl.1-Abl.3}}, replacing AuEmo Tokenizer with alternative emotion representations leads to clear degradation in emotional expressiveness. For example, AuEmoChat improves AuEmoACC from 20.38 in Abl.1 to 28.71. This indicates that the learned authentic emotion token space is more effective than limited emotion labels, open-vocabulary emotion labels, and emotion captions, enabling the model to better capture emotional states in conversational speech. \textbf{\textit{In Abl.4-Abl.5}}, removing AuEmoToMe or its AuEmo-guided merging strategy degrades both speech quality and emotional expressiveness. For example, compared with Abl.4, AuEmoChat improves N-DMOS from 3.817 to 4.171 and EmoACC from 57.97 to 61.04. This suggests that redundant dialogue tokens hinder effective context modeling, whereas AuEmoToMe improves target emotion understanding and speech token prediction by reducing irrelevant information. \textbf{\textit{In Abl.6-Abl.8}}, removing the AuEmo condition, AuEmo classifier guidance, or merged dialogue context degrades both speech quality and emotional expressiveness. This demonstrates that jointly modeling authentic emotion and merged dialogue context is essential for generating context-consistent and emotionally expressive speech. Notably, Abl.1, which uses traditional limited emotion labels, achieves the worst performance on both E-DMOS and AuEmoACC, highlighting the importance of modeling authentic emotion. In Abl.8, removing the merged dialogue context in flow matching significantly degrades speech quality. For example, WER increases from 9.14 to 13.63, MCD increases from 6.847 to 7.610, and SpkSIM decreases from 78.03 to 72.61. These results indicate that contextual information plays a critical role in speech generation.

\begin{table}[t]
  \centering
  \caption{Analysis results of key design choices in AuEmoCodec. \colorbox{blue_1}{Blue} indicates the best performance, and \colorbox{pink_1}{pink} indicates the second-best performance.}
  % \vspace{-2mm}
  \label{tab:auemocodec_any}
  \small
  \setlength{\tabcolsep}{3.5pt} 
  \resizebox{0.98\columnwidth}{!}{
  \begin{tabular}{l | ccccc} 
    \toprule
    \textbf{Methods} & \textbf{Tol@1} & \textbf{Tol@2}  & \textbf{S-EAA} & \textbf{D-EAA} & \textbf{Usage} \\
    \midrule
    
    \rowcolor{gary_1}
    \multicolumn{6}{l}{\textit{\textbf{Comparative Study of Quantization Paradigms}}} \\ 
    \arrayrulecolor{gray!50} \hline 
    VQ  & 35.34 & 62.55  & 10.85 & 47.46  & 55.10\% \\
    RVQ  & 54.39 & 81.07  & \cellcolor{pink_1}18.53 & \cellcolor{pink_1}71.25  & 61.10\% \\
    R-FSQ  & 53.97 & \cellcolor{pink_1}81.69  & 16.25 & 66.15  & \cellcolor{blue_1}87.30\% \\
    G-FSQ  & \cellcolor{pink_1}54.46 & \cellcolor{blue_1}\textbf{82.24}  & 18.30 & 70.81  & 72.10\% \\
    \textbf{FSQ}      & \cellcolor{blue_1}\textbf{55.04} & 81.17 & \cellcolor{blue_1}\textbf{18.61} & \cellcolor{blue_1}\textbf{71.77}  & \cellcolor{pink_1}75.00\% \\
    \midrule
    
    \rowcolor{gary_1}
    \multicolumn{6}{l}{\textit{\textbf{Comparative Study of Codebook Capacity}}} \\
    \arrayrulecolor{gray!50} \hline 
    Codebook Size (200)                     & 45.82 & 79.52 & 10.27 & 46.44  & \cellcolor{pink_1}76.00\% \\
    Codebook Size (600)                    & 46.62 & 79.75 & 10.01 & 46.96  & \cellcolor{blue_1}83.16\% \\
    \textbf{Codebook Size (1000)}      & \cellcolor{blue_1}\textbf{55.04} & \cellcolor{pink_1}81.17 & \cellcolor{blue_1}\textbf{18.61} & \cellcolor{blue_1}\textbf{71.77}  & 75.00\% \\
    Codebook Size (1400)                    & 52.26 & 81.00 & 17.21 & 70.50 & 67.28\% \\
    Codebook Size (2000)                  & 52.82 & 80.99 & 17.73 & \cellcolor{pink_1}70.59 & 52.10\% \\
    Codebook Size (4000)                  & \cellcolor{pink_1}54.28 & \cellcolor{blue_1}\textbf{81.21} & \cellcolor{pink_1}18.30 & 70.30  & 41.42\% \\

    \midrule
    \rowcolor{gary_1}
    \multicolumn{6}{l}{\textit{\textbf{Comparative Study of Different LLM-as-a-Judge Models}}} \\ 
    \arrayrulecolor{gray!50} \hline 
   
    Qwen2-Audio & 39.99 & 74.12 & 15.17 & 54.44 & 73.40\% \\
    Kimi-Audio & 43.25 & 78.54 & 17.25 & 64.07  & 74.10\% \\

    Qwen3-Omni-Captioner & 43.04 & \cellcolor{blue_1}\textbf{81.51} & 17.88 & 68.51 & \cellcolor{blue_1}\textbf{75.90\%} \\
     \textbf{Gemini-2.5-Flash}   & \cellcolor{blue_1}\textbf{55.04} & \cellcolor{pink_1}81.17 & \cellcolor{blue_1}\textbf{18.61} & \cellcolor{blue_1}\textbf{71.77}  & \cellcolor{pink_1}75.00\% \\
    \bottomrule
  \end{tabular}
  }
\end{table}

\subsection{Analysis of AuEmoCodec Architecture}
In this section, we analyze the key design choices of AuEmoCodec from three aspects: quantization paradigms, codebook capacity, and the judge model, as shown in Table \ref{tab:auemocodec_any}. \textbf{\textit{For quantization paradigms}}, we evaluate Vector Quantization (VQ) \cite{van2017neural}, Residual VQ (RVQ), Finite Scalar Quantization (FSQ) \cite{mentzer2023finite}, Residual FSQ (R-FSQ), and Grouped FSQ (G-FSQ). FSQ achieves the best performance on Tol@1, S-EAA, and D-EAA, and shows the most stable overall performance. Although its codebook usage is slightly lower than R-FSQ, it achieves stronger authentic emotion modeling, indicating that FSQ provides a better balance between representation capacity and efficiency. \textbf{\textit{For codebook capacity}}, as the codebook size increases from 200 to 1000, the performance on Tol@1, S-EAA, and D-EAA consistently improves, indicating that a larger codebook enhances the discrimination ability and representation precision of authentic emotions. However, as the codebook size increases further, codebook usage decreases and performance degrades, suggesting that an excessively large codebook reduces effective utilization and impacts authentic emotion modeling performance. Overall, a codebook size of 1000 achieves the best balance between performance and usage. \textbf{\textit{For the judge model selection}}, we evaluate Qwen2-Audio \cite{chu2024qwen2}, Kimi-Audio \cite{ding2025kimi}, Qwen3-Omni-Captioner \cite{xu2025qwen3}, and Gemini-2.5-Flash \cite{comanici2025gemini}. Gemini-2.5-Flash achieves the best performance on Tol@1, S-EAA, and D-EAA, demonstrating stronger authentic emotion perception ability. In contrast, Qwen3-Omni-Captioner performs better on Tol@2 and usage. Overall, Gemini-2.5-Flash provides more accurate emotion supervision, which helps learn high-quality emotion representations \cite{luo2025instructionquality}. Based on the above analysis, we adopt FSQ as the quantization method, set the codebook size to 1000, and use Gemini-2.5-Flash as the judge model.

\begin{figure}
    \centering
    \includegraphics[width=1\linewidth]{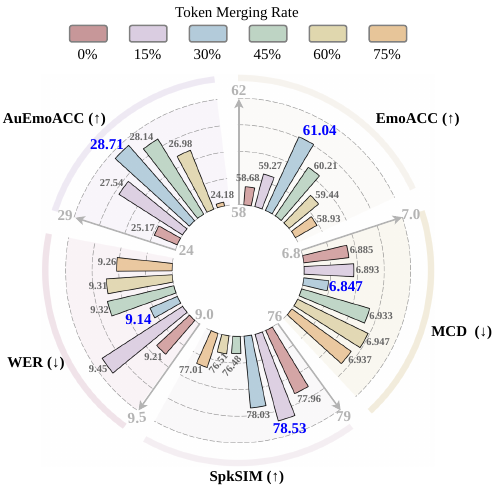}
    \caption{Analysis results of different token merging rates on speech quality and emotion expressiveness.}
    % \vspace{-4mm}
    \label{fig:radio}
\end{figure}

\subsection{Analysis of Context Token Merging Rate}
To analyze the impact of the token merging rate in AuEmoToMe, we conduct experiments with merging ratios of 0\%, 15\%, 30\%, 45\%, 60\%, and 75\%. As shown in Fig. \ref{fig:radio}, we report the performance on AuEmoACC, EmoACC, WER, MCD, and SpkSIM, where the five sectors correspond to different evaluation metrics. Overall, a token merging rate of 30\% achieves the best overall performance. Under this setting, the model attains the best results on AuEmoACC, EmoACC, WER, and MCD, while maintaining near-best performance on SpkSIM. \textbf{\textit{For emotion metrics}}, as the merging ratio increases from 0\% to 30\%, both AuEmoACC and EmoACC consistently improve. This indicates that moderate token merging effectively reduces the interference of redundant tokens and enhances the model’s ability to capture the authentic emotion of the target utterance. When the merging ratio is further increased, the performance degrades, suggesting that excessive merging impairs emotional expressiveness. \textbf{\textit{For speech quality metrics}}, WER and MCD achieve the best performance at 30\%, indicating that appropriate token merging alleviates the interference of redundant context in speech modeling, thereby improving pronunciation accuracy. At higher merging ratios, the performance degrades, further confirming the negative impact of information loss on speech generation. Finally, we adopt a token merging rate of 30\% during inference to achieve the best balance between emotional expressiveness and speech quality.

\section{Conclusion and Future Work}
In this work, we propose AuEmoChat, a novel CSS framework for authentic emotion understanding and rendering. AuEmoChat introduces AuEmoCodec to learn an authentic emotion token space, AuEmoToMe to merge redundant multimodal context tokens while preserving emotion context information, and Authentic Emotion Flow Matching to generate context-consistent authentic emotional speech. To the best of our knowledge, AuEmoChat is the first CSS system explicitly devoted to authentic emotion modeling. We hope this work can inspire further research on more human-like emotional speech synthesis for conversational agents. In future work, we will extend AuEmoChat to multilingual settings, scale up AuEmoCodec training, and further improve the interpretability of authentic emotion for more comprehensive emotional CSS. More detailed limitations and future works are provided in Appendix D.

\section{Acknowledgments}
This research of Rui Liu was funded by the General Program (No. 62476146) of the National Natural Science Foundation of China, the Young Elite Scientists Sponsorship Program by CAST (No. 2024QNRC001), the Outstanding Youth Project of Inner Mongolia Natural Science Foundation (No. 2025JQ011), the Key R\&D and Achievement Transformation Program of Inner Mongolia Autonomous Region (No.2025YFHH0014), and the Central Government Fund for Promoting Local Scientific and Technological Development (No. 2025ZY0143). The work of Haizhou Li was supported by the Shenzhen Science and Technology Research Fund (Fundamental Research Key Project, Grant No.
JCYJ20220818103001002) and the Program for Guangdong Introducing Innovative and Entrepreneurial Teams (Grant No. 2023ZT10X044).

%%
%% The next two lines define the bibliography style to be used, and
%% the bibliography file.
\bibliographystyle{ACM-Reference-Format}
\bibliography{auemo}

%%
%% If your work has an appendix, this is the place to put it.

\end{document}